\documentclass{ws-ijmpcs}
\newcommand       \citep          {\cite}
\newcommand       \citet          {Ref.~\refcite}

\newcommand       \aap          {A\&A }
\newcommand       \ssr           {Spac.Scienc.Review}

\newcommand       \nat         {Nature}
\newcommand       \araa        {ARA\&A}
\newcommand       \pasj        {}

\newcommand       \apj        {ApJ}
\newcommand       \apjl        {ApJL}
\newcommand       \mnras   {MNRAS}
\newcommand       \memsai {MemSAIt}

\begin{document}

\markboth{Niccol\`o Bucciantini}
{Pulsar Wind Nebulae Modeling}

%
\catchline{}{}{}{}{}
%

\title{PULSAR WIND NEBULAE MODELING
}

\author{NICCOL\`O BUCCIANTINI}

\address{INAF Osservatorio Astrofisico di Arcetri\\
L.go Fermi 5, Firenze, FI, 50125,
Italy\\
niccolo@arcetri.astro.it}

\maketitle

\begin{history}
\received{Day Month Year}
\revised{Day Month Year}
\end{history}

\begin{abstract}
Pulsar Wind Nebulae (PWNe) are ideal astrophysical laboratories where high energy relativistic
phenomena can be investigated. They are close, well resolved in our
observations, and the knowledge
 derived in their study has a strong impact in many other fields, from AGNs to GRBs. Yet
there are still unresolved issues, that prevent us from a full clear understanding of these objects.
The lucky combination of high resolution X-ray imaging and numerical codes to handle the outflow
and dynamical properties of relativistic MHD, has opened a new avenue of investigation that has
lead to interesting progresses in the last years. Despite all of this, we do not understand yet
how particles are accelerated, and the functioning of the pulsar wind and pulsar magnetosphere,
that power PWNe. I will review what is now commonly known as the MHD
paradigm, and in particular I will focus on various approaches that
have been and are currently used to model these systems. For each I
will highlight its advantages, limitations, and degree of applicability.

\end{abstract}

\ccode{PACS numbers: 98.38.Mz, 97.60.Gb}

\section{Introduction}

Pulsar Wind Nebulae (PWNe) are bubbles of relativistic particles,
pairs and perhaps ions. They are due to the interaction of the ultra-relativistic wind
from a pulsar either with the parent
Supernova Remnant (SNR) or the
ISM\citep{Chevalier04a,Gaensler_Slane06a,Bucciantini08a}.
They emit a broad band spectrum extending from radio to $\gamma$-rays, via
synchrotron and Inverse Compton emission\citep{Weiler_Panagia78a,Asaoka_Koyama90a,Harding96a,Bandiera00a,Kargaltsev_Pavlov10a}. Among
all PWNe, the
Crab Nebula\citep{Hester08a} takes a special place. It is such a well
known system, that its properties are a benchmark for our models.   

To model a PWN one needs an
acceleration mechanism responsible for producing the non-thermal
particle's distribution function that is observed, and a dynamical model describing the advection
and evolution of the flow.

To these days the acceleration problem remains unsolved. Ions in the pulsar wind, carrying a large
fraction of the total wind energy, could accelerate pairs, but there is no compelling
evidence for their presence\citep{Amato_Arons06a,Sironi_Spitkovsky11b}. Efficient {\it
  Fermi-like} acceleration of a pure pair plasma, requires a wind with
an eccessively small magnetization\citep{Sironi_Spitkovsky09a}. Reconnection of a striped
wind\citep{Lyubarsky03a,Lyubarsky_Livers08a,Sironi_Spitkovsky11a},
leads to hard spectra, as the ones observed in radio, but
requires winds with properties not compatible with standard pulsar wind dynamics\citep{Goldreich_Julian69a,Bogovalov97a,Hibschman_Arons01a}. There is also the
possibility of continuous distributed turbulent acceleration in the bulk of the
nebula\citep{Barnes_Scargle73a,Bucciantini_Arons+11a}. There is however a general consensus that X-ray emitting particles are
accelerated at the wind termination shock. The origin of radio
emitting ones is still currently debated\citep{Atonyan_Aharonian96a}.

This contrasts with dynamical models for the flow properties  that
have been
developed to a greater success. They are based now on a MHD
paradigm. In fact the gyro-radius
of the  particles is much smaller that the
size of these  nebulae, even up to the highest
observed energies. A fluid description is thus possible, and can be
formulated in terms of MHD.  To build an emission model, one just
needs to couple it with a prescription for
particle's acceleration. The hope is to reproduce
the observed morphology, to relate it  to the pulsar wind
properties, and to provide a unifying picture for all these objects.

Recently, the tools and approches that have been used in the
investigation of PWNe have been applied to Long and Short
GRBs\citep{Dall'Osso-Stratta+11a,Bucciantini-Quataert+08a,Bucciantini-Quataert+09a,Bucciantini-Metzger+12a}. Indeed these systems too can be modeled as expanding
relativistic magnetized bubbles, and share a large amount of physics
and dynamics with PWNe. 

I will review in the following the various approaches that have been
presented in the literature to model these systems, discussing their
pros and cons, in relation to the questions one would like to answer.

\section{1-Zone Models}

We call {\it 1-Zone  model} any model where the
PWN is described as an uniform object, with a given size (radius)
possibly time-dependent. The internal structure is completely neglected. These models, notwistanding their simplicity,  have proved very successful
for the investigation of the global average properties of PWNe. In particular their interaction with the
partent SNR. The first 1-Zone model for PWNe was developed by \citet{Pacini_Salvati73a}, with application to the Crab nebula. They
studied  the evolution of particles in an expanding bubble, subject to adiabatic losses,
radiation losses and a continuous injection. The analytical approach
they adopted is however justifiable only
for young systems in the so called {\it free-expansion phase} (the first few
thousands years). Subsequently this model has been extended to take partially into account the interaction with the
SNR\citep{Reynolds_Chevalier84a,Reynolds_Chanan84a}, or changes in
injection history\citep{Atoyan99a}. Only recently a numerical
approach was been used\citep{Bednarek_Bartosik03a,Torres_Cillis+13a,Tanaka_Takahara13a,Tanaka_Takahara10a,Matin_Torres+12a,Gelfand_Slane+09a,Bucciantini_Arons+11a}. This
allows us to follow the evolution for a longer time and to take into
account the various phases of the interaction with the parent SNR. 
Moreover we can model  many different distribution functions,
that are not necessarily a single power law (either broken power laws
have been considered, or a power-law plus a Maxwellian).  In
particular, the work by \citet{Gelfand_Slane+09a} is exemplary for
showing the richness of behaviors that is found and the degree of
complexity that one can investigate using these simple tools.

1-Zone models are quite easy to build and they allow us to sample a large
parameter space of often unknown injection or evolutionary
properties.  The Radio
emission properties, the location of spectral breaks, the ratio of
synchrotron to Inverse Compton emission, can be fairly well
modelled. Of course one should not expect more than a rough
quantitative agreement with the global observed properties of these
systems, and they are not reliable for the high-energy
X-ray properties, that depend on the post shock flow.
They are also quite good in recovering the
duration and timescale of the various phases of interaction with the
SNR.

Unfortunately, quite often the quality of the available data is
limited, and
1-Zone models are then the only justifiable approach. They
also constitute a good starting point for more accurate,
multidimensional investigations.

\section{1-D Models}

It is a general expectation that, increasing the dimensionality of a
model, should lead to a more realistic description of a system: a
description closer to the true working. Including a dependence on
the radial coordinate from the
pulsar leads to {\it 1-D models}.  The first 1-D model of the Crab
Nebula was presented by \citet{Rees_Gunn73a}. Several spatial features
 could be explained.
\begin{itemize}
\item the under-luminous region, centered on the location of
the pulsar, and interpreted as the ultra-relativistic unshocked
wind. 
\item the wavelength dependent size of the Crab Nebula, with radio
emission extending beyond the X-ray one,  explained in term
of combined bulk flow from the termination shock to the outer edge of
the nebula, and synchrotron cooling. 
\item the existence of a bright torus, understood as due to compression of the magnetic field in the outer
layers of the nebula. 
\end{itemize}
This was also the fist time nebular properties were used to infer the
Lorentz factor and/or magnetization of the wind. This model has
been further developed by \citet{Kennel_Coroniti84a,Kennel_Coroniti84b} who solved the full set of
relativistic MHD equations, and provided a quantitative estimate of
the structural properties of the Crab Nebula. Later \citet{Emmering_Chevalier87a}
presented a time dependent analytical solution. 

More recently numerical
investigations have allowed us to avoid many of the simplifications
required by the early analytical
treatment. \citet{Blondin-Chevalier+01a} and \citet{van_der_Swaluw-Achterberg+01a} were the first to investigate the dynamics
of the PWN-SNR interaction using a hydrodynamical code, for all the
various evolutionary
phases: the {\it
  free-expansion phase}\citep{Chevalier-Fransson92a}, the {\it reverberation phase}, and
the final {\it Sedov phase}. Similar works have also been carried out 
by \citet{Bucciantini-Blondin+03a,Bucciantini-Bandiera+04a} which were the first to use relativistic
MHD and to investigate the role of magnetic field. More recently
 \citet{de_Jager-Fereira+08a} have used a 1-D model to compute the evolution of an
injected distribution function, and to compute an integrated spectrum
for G21.5-0.9.

Out of 1-D models was born the so called {\it $\sigma$
  problem}\citep{Melatos98a}. The  {\it $\sigma$
  problem} is two-faced: one related to the
efficiency of accelerating particles at a shock, the other to the
ability of slowing down a relativistic magnetized outflow. Here I will only
consider this latter
aspect. 
In Ideal MHD, a radial highly-magnetized (with a toroidal field) supersonic
relativistic pulsar wind cannot be slowed down by a shock to match
the typical expansion speeds of the confining SNR. It only happens for
a small value\citep{Kennel_Coroniti84a}  of the magnetization $\sim 10^{-3}$. This contrasts strongly the
expectations of pulsar wind electrodynamics\citep{Michel69a,Goldreich-Julian70a,Beskin-Kuznetsova+98a,Bogovalov01a}, which suggest that the
magnetization in the wind, at best, with minor deviations from radial outflow, could be $\sim 1$ at the typical
distance of the nebulae\citep{Chiueh_Li+98a}. 

Moreover 1-D models fail to explain
the relative size of X-ray an Radio in the Crab Nebula\citep{Amato-Salvati+00a}, and the correct softening of the spectrum
with the distance from the pulsar\citep{Reynolds04a}. These problems
are related to the fact that in 1-D models there is a one-to-one
correspondence between distance, age, magnetic field and flow
velocity. This inevitably leads to onion-like structures, which are
the exact opposite of the complete mixing at the base of 1-Zone models.

\section{2-D Models}

The interest in multidimensional models for PWNe grew mostly in the
past ten years, and was mainly driven by optical and X-ray
images from HST, CHANDRA and XMM-Newton. High resolution images have shown that the inner
region of PWNe is characterized by a complex axisymmetric structure,
generally referred as {\it jet-torus structure}. It was observed at first in the
Crab Nebula\citep{Hester-Scowen+95a,Weisskopf-Hester+00a}, and it has since been
detected in many other PWNe\citep{Gaensler-Arons+02a,Lu-Wang+02a,Camilo-Gaensler+04a,Pavlov-Teter+03a,Romani-Ng03a,Romani-Ng+05a,Slane-Helfand+04a}.
Today there is
a general consensus that such structure is always present in young
nebulae.

It is evident that, in the presence of a toroidal magnetic field,
anisotropic stresses will arise in the nebula, defining a preferential
symmetry axis. Models that assume a cylindrical geometry
are referred as {\it 2-D models}. \citet{Begelman-Li92a} were the first to develop a 2-D model accounting
for magnetic tension, in an attempt to explain the observed prolate
geometry of the Crab Nebula, and to use it as a further constrain on
the plasma magnetization. 

Given the complex dynamics, associated with the
larger degrees of freedom due to the increased dimensionality, in
order  to
move beyond the simple analytical solution by
 \citet{Begelman-Li92a}, we had to wait for numerical codes for
computational fluid dynamics.\citep{Komissarov99a,Del_Zanna-Bucciantini+03a}
 \citet{van_der_Swaluw03a} was the first to present a numerical model
of a prolate nebula applied to 3C58. It was however the investigation
of the dynamics in the vicinity of the termination shock that drove most of the successive
attempts at modeling the inner jet-torus structure. 

At the typical distance of the termination shock,
the wind luminosity and magnetization are not uniform, but can be
described by the so called {\it split monopole} solution: most of the
energy is confined to the equatorial region, and there is no evidence
of jets or collimated pulsar outflows along the axis. This is
confirmed both by theoretical\citep{Begelma-Li94a,Beskin-Kuznetsova+98a}
and numerical\citep{Contopoulos-Kazanas+99a,Bogovalov01a,Timokhin06a,Komissarov06a,Spitkovsky06a,Bucciantini-Thompson+06a}
studies. Such a wind would naturally drive a complex dynamics at least in
the inner region. \citet{Bogovalov-Khangoulian02a,Bogovalov-Khangoulian02b} were the first to investigate this
problem, modeling the flow dynamics at the termination shock. Later
 \citet{Lyubarsky02a} suggested that hoop stresses in the body of the nebula
could lead to the collimation of the jet. This has been extensively confirmed
numerically by different groups\citep{Komissarov-Lyubarsky04a,Del_Zanna-Amato+04a,Bogovalov-Chechetkin+05a,Del_Zanna-Volpi+06a,Volpi-Del_Zanna+08a}. The
results of this numerical investigation went far beyond the
confirmation of these simple ideas. Many aspects of the nebular
morphology have been investigated: the details of emission maps\citep{Del_Zanna-Volpi+06a},  polarization\citep{Bucciantini-Del_Zanna+05a}, the integrated broad-band spectrum\citep{Volpi-Del_Zanna+08a}. We can now reproduce reasonably well many
of the observed X-ray features of the Crab Nebula, including
the wisps and the knot, in terms of relative size and
brightness. Recently the same tools have been applied to the modeling
of variability. PWNe have been known to be variable in the X-rays since the first
observations, in particular in the wisps region\citep{Hester-Scowen+95a,Schweizer-Bucciantini+13a}. Such variability
can be reproduced by 2-D models\citep{Bucciantini-Del_Zanna06a,Bucciantini08a,Camus-Komissarov+09a}
both in term of its typical timescale and in its morphological
pattern, of outgoing waves.

Concerning the wind magnetization $\sigma$,  2-D models have shown
that values $\sim 0.03-0.1$ are required to
explain the observed morphology and the presence of a jet; about
one order of magnitude higher than in 1-D models. This for two reasons:
the extra degree of freedom of the 2-D geometry allows to accommodate more
magnetized winds; a large degree of internal fluid
turbulence and mixing. This strongly suggest that reality could be
closer  to the full mixing
assumption of 1-Zone models, instead of
the onion-like structure of 1-D models.

Mostly developed to investigate the dynamics of the internal region, 2-D models have been
also applied to the evolution of the PWN-SNR system. \citet{Blondin-Chevalier+01a} were the first to investigate the 2-D dynamics
of the reverberation phase, showing the high level of mixing between
the SNR and PWN, and the role of density gradients in the ISM, leading
to a relic PWN displaced with respect to the pulsar, as it is often
observed in old systems. \citet{Jun98a} and  \citet{Bucciantini-Amato+04a} were the
first to investigate the Rayleigh-Taylor instability at the interface
with the SNR ejecta, both in the magnetized an unmagnetized case,
showing the development of the filamentary network observed in the
Crab Nebula. \citet{van_der_Swaluw-Downes+04a} investigated in detail the
interaction of a pulsar moving across the SNR, the transition from
spherical PWNe in the early phase to a cometary structure, and in
 \citet{van_der_Swaluw-Achterberg+02a} the interaction with the SNR forward shock was also taken
into account. Recently \citet{Ferreira-De_Jager08a} analyzed the role of ISM density gradients,
in the presence of an ordered magnetic field.  

\section{3-D Models}

It has been known for a long time\citep{Begelman98a,Nalewajko-Begelman12a} that 2-D configurations with a
purely toroidal magnetic  field, in 3-D are subject to current driven instability ({\it
  kink instability}). Instability of the jet is seen in PWNe\citep{Pavlov-Teter+03a,Mori-Burrows+04a}.
To fully capture the dynamics associated with
this instability full {\it 3-D models} are necessary. Only recently this process has been
studied with numerical codes in the simplified regime of a
magnetized column confined by a hot atmosphere\citep{O'Neill-Beckwith+12a,Mizuno-Lyubarsky+11a,Mizuno-Lyubarsky+12a}. The instability reaches the non-linear regime on a typical
Alfv\`enic timescale, and after about 10 Alfv\`enic timescales, the
original toroidal field has almost completely vanished. This process
could easily solve the $\sigma$ problem\citep{Begelman98a}, but one
must be carefull because the high level of polarization\citep{Smith-Jones+88a,Velusamy85a,Bietenholz91a} suggests that this
instability cannot be so violent. Part of the reason for this
inconsistency can be found in the simplified regime that these models
consider. In PWNe the toroidal magnetic field is continuously injected by
the pulsar wind, and the PWN is not pressure confined but confined by a
high density wall at the contact discontinuity with the SNR ejecta. This
implies that there must be a balance between the
injection and dissipation of the toroidal field.  Moreover  in the inner region, close to the termination shock, where the
plasma is injected, the magnetic field
is bound to be close to toroidal, because instabilities would preferentially act in
the outer regions. Unfortunately a full 3-D simulation of a PWN,
evolving long
enough to reach a balance, has not been carried out yet, mostly
because of computational cost.  

However recently \citet{Porth-Komissarov+13a} (see also these
proceedings for a more detailed discussion) presented a very interesting preliminary
study, where the dynamics of the PWN in a full 3-D setting was
investigated for a time-length corresponding to 70 yr. While this is
about one order of magnitude smaller than the age of the Crab Nebula, and it
can be debated if the nebula has reached a state of dynamical
equilibrium, nevertheless this timescale is longer that a typical
Alfv\`enic crossing time, and longer than the synchrotron cooling time for
X-ray emitting particles. The results confirm the qualitative expectations: the
inner region close to the termination shock is still dominated by a toroidal field, and
preserves the axisymmetric structure of 2D simulations; a jet is formed
and extends in the body of the nebula; instabilities at the outer edge of
the nebula act to reduce the net toroidal field component. It is also
found that the nebula can accommodate even a wind with high $\sigma$, of order
unity. Recalling that, even minor deviations of the wind from a purely radial
flow, can lower $\sigma$ to values of order unity, at the typical
distance of the termination shock radius\citep{Chiueh_Li+98a}, this suggests that the $\sigma$ problem is likely an artifact of the
1-D geometry of theories of old. However this does not rule out the
possibility of other non-ideal effects, like dissipation and/or
reconnection to play a role. It also does not solve the $\sigma$
problem from the point of view of particle's acceleration. But it
suggests that one could seek for independent solutions of these two
aspects, instead of looking for one that solves both at once.

\section{Bow-Shocks}

Despite the fact that few works have been devoted to
the multi-dimensional investigation of the dynamical evolution of old
PWNe inside  SNRs, there are several works focusing on
the interaction with the ISM. Pulsars moving in the ISM produce the so
called {\it pulsar wind bow-shock nebulae}\citep{Cordes-Romani+93a,Bell-Bailes+95a,Gaensler-van_der_Swaluw+04a,Pavlov-Sanwal+06a,Kargaltsev-Misanovic+08a,Ng-Bucciantini+12a}. The
first attempt to model the physics
 of these
systems, accounting for the presence of neutral Hydrogen in the ISM, was done by \citet{Bucciantini-Bandiera01a,Bucciantini02b} extending the {\it thin-layer
  approximation}\citep{Giuliani82a} used to model cometary nebulae
\citep{Bandiera93a,Wilkin96a,Wilkin00a}. The {\it thin layer}
approximation neglects the thickness of the nebula. However these
models provide a good description of the shape and physical properties (thickness to hydrogen penetration,
H$_\alpha$ luminosity) in 
the head region. This was confirmed by more accurate 2-D
axisymmetric simulations both in the hydrodynamical regime\citep{Bucciantini02a,Gaensler-van_der_Swaluw+04a}
and in the relativistic MHD regime\citep{Bucciantini-Amato+05a}. A 3-D extension of the study of these
systems to take into account either a non-uniform ambient medium, or the
anisotropy in the pulsar wind energy flux (in analogy with 2-D
simulations of Crab-like PWNe) has been presented by \citet{Vigelius-Melatos+07a}.

Bow-shock models have also been recently developed and applied to the
interaction and confinement of the pulsar wind
in binary systems\citep{Bogovalov-Khangulyan+08a,Bogovalov-Khangulyan+12a,Bosch-Ramon-Barkov+12a}. In
this case it is the wind from the companion that provides the confining
medium. These models have been used to explain orbital variations and modulations of the
high energy comptonized emission.




\end{document}